\documentclass[pra,aps,twocolumn,showpacs,amsmath,amssymb]{revtex4}
\usepackage{epsfig}
\usepackage{graphics}
\usepackage{bm}
\usepackage{amssymb,amsmath}
\usepackage{graphicx}

\newcommand{\R}{\mathcal{R}}

\newcommand{\e}{\mathrm{e}}                  
\newcommand{\ket}[1]{|#1\rangle}             
\newcommand{\bra}[1]{\langle #1|}            
\newcommand{\braket}[2]{\langle #1|#2\rangle}
\renewcommand{\d}[1]{\mathrm{d}#1}           
\newcommand{\tr}{\mathrm{Tr}}                
\renewcommand{\Re}{\mathrm{Re}}              
\newcommand{\mt}[1]{\mathrm{#1}}             
\newcommand{\la}{\langle}
\newcommand{\ra}{\rangle}
\newcommand{\wh}[1]{\widehat{#1}}
\newcommand{\be}{\begin{equation}}
\newcommand{\ee}{\end{equation}}
\newcommand{\beqa}{\begin{eqnarray}}
\newcommand{\eeqa}{\end{eqnarray}}
\newcommand{\beq}{\begin{equation}}
\newcommand{\eeq}{\end{equation}}

\begin{document}
\title{Relation between quantum dwell times and flux-flux correlations}
\author{J. Mu\~noz}
\email{josemunoz@saitec.es}
\author{D. Seidel}
\email{dirk_seidel@ehu.es}
\author{J. G. Muga}
\email{jg.muga@ehu.es}

\address{Departamento de Qu\'{\i}mica-F\'{\i}sica, UPV-EHU, Apartado Postal 644, 48080 Bilbao, Spain}

\begin{abstract}
We examine the connection between the dwell time of a quantum particle 
in a region of space
and flux-flux correlations at the boundaries. 
It is shown that the first and second moments of a flux-flux correlation function 
which generalizes a previous proposal by Pollak and Miller [E. Pollak and W. H. Miller, 
Phys. Rev. Lett. {\bf 53}, 115 (1984)],  
agree with the corresponding moments of the dwell-time distribution,
whereas the third and higher moments do not. 
We also discuss operational approaches and approximations to measure the
flux-flux correlation function and thus the second moment of the dwell time,
which is shown to be characteristically quantum, and larger than the corresponding classical moment even for freely moving 
particles.       

\end{abstract}
\maketitle

\section{Introduction}
Time observables, i.e., times measured as random variables 
after a microscopic system is prepared, e.g. according to a 
given wavefunction, 
are very common in the laboratories. 
Examples are the lifetimes of excited species, or the arrival
times of ions, 
atoms or molecules at a scintillation detector, a microchannel plate or a laser-illuminated region.  
Nevertheless,  
incorporating time observables into quantum mechanics is a problematic 
task \cite{Muga-book}, and it has even been discouraged by many 
physicists influenced by Pauli's theorem
and the extended notion that ``time is only a parameter'' within the quantum realm.  
Time observables are characterized
experimentally, for a given  preparation of the initial state,
by a distribution or by its 
statistical moments, and useful information may be extracted from them.
The spectacular progress in quantum state
manipulation with laser and magnetic 
cooling techniques emphasizes the need to treat atomic motion quantally rather
than classically \cite{MuLea-PR-2000,DEHM2002}, 
and therefore a quantum approach to time quantities. 
Much work has been carried out in the last two decades 
trying to formulate time observables theoretically 
and also to connect the abstract proposals with actual or idealized experiments. 
Central to these investigations have been the tunneling time,
the arrival time, and the dwell time \cite{Muga-book}.    

The dwell time of a particle in a region of space and its close relative, 
the delay time \cite{Smith60}, are in particular rather fundamental quantities 
that characterize the duration
of collision processes, the lifetime of unstable systems, the response to perturbations, or the properties of chaotic scattering. 
In addition, the importance of dwell and delay times is underlined by 
their relation to the density of states, 
and to the virial expansion in statistical mechanics \cite{Nuss2002}.

The theory of the quantum dwell time is  
quite peculiar and subtle in several respects \cite{HauSto-RMP-1989, LaMa-RMP-1994}.
To begin with, unlike other time quantities, there has been 
a broad consensus on its operator representation \cite{EkSie-AP-1971, JaWa-PRA-1989},
\beq
\wh{T}_D = \int_{-\infty}^\infty \d t\,\wh{\chi}_\R(t),
\eeq
where $\wh{\chi}_\R(t)$ is the (Heisenberg) projector onto a region of space,
which we shall limit here to one dimension for simplicity,
$\R = \{x: x_1 \leq x \leq x_2\}$,   
\beq
\wh{\chi}_\R(t)=e^{i\wh{H}t/\hbar}\int_{x_1}^{x_2} dx
|x\ra \la x| e^{-i\wh{H}t/\hbar}.
\eeq
We shall also assume that the Hamiltonian holds a purely continuous spectrum
with degenerate (delta-normalized) scattering eigenfunctions $|\phi_{\pm k}\ra$ corresponding to 
incident plane waves $|\pm k\ra$, with energy $E=k^2\hbar^2/(2m)$.  
The concept of an average dwell time $\tau_D(k)$ for a finite space region  in the stationary regime is due to B\"uttiker \cite{Bue-PRB-1983},
\be \label{eq:dwell_stat}
\tau_D(k) = \frac{1}{j(k)} \int_{x_1}^{x_2} \d x\, |\phi_k(x)|^2,
\ee
where $j(k)$ is the incoming flux associated with $\ket{\phi_k}$. 
 
The operator $\wh{T}_D$ is positive definite and essentially selfadjoint.  Moreover, being a ``time duration'' rather than a time instant, $\wh{T}_D$ commutes with the Hamiltonian without conflict with Pauli's theorem, and therefore it can be diagonalized in the eigenspace of $\wh{H}$. 
This simplifies the derivation of the corresponding quantum dwell-time distribution, which, for a state $|\psi\ra=|\psi(t=0)\ra$ is formally given by
\be \label{eq:dwell_distri}
\Pi(\tau) = \bra{\psi} \delta(\wh{T}_D -\tau) \ket{\psi}.
\ee
Following the same manipulation done for the $S$ operator in one-dimensional 
scattering theory,  
it is convenient to define an on-the-energy-shell
$2\times 2$ dwell-time matrix $\mathsf{T}$, by factoring out an energy delta,  
\beq
\la \phi_k|\wh{T}_D|\phi_k'\ra=\delta(E_k-E_{k'})\frac{|k|\hbar^2}{m}
\mathsf{T}_{kk'}, 
\label{dele} 
\eeq
where $\la \phi_k|\phi_{k'}\ra=\delta(k-k')$ and  
\beq
\mathsf{T}_{kk'}=\la \phi_k|\wh{\chi}_\R|\phi_{k'}\ra\frac{hm}{|k|\hbar^2},\;\; 
E_k=E_{k'}
\eeq
In particular, $\tau_D(k)=\mathsf{T}_{kk}$.
Diagonalization of $\mathsf{T}$ for free motion leads generically to an 
interesting quantum
peculiarity \cite{DEMN04}:
the existence of two different dwell-time eigenvalues
$t_\pm(k) = mL[1\pm \sin(kL)/kL]/\hbar k$ on the energy shell, $L=x_2-x_1$, 
at variance with the classical case. As a consequence, a broader
spread (variance) in the quantum distribution than the one 
expected for a corresponding ensemble of classical particles will occur.           
   
In spite of the nice properties of $\wh{T}_D$,  a 
direct and sufficiently non-invasive measurement of the dwell time of a 
quantum particle 
in a region of space, so that the statistical moments are produced by averaging over  
individual dwell-time values, is yet to be discovered. If the particle is detected (and thus localized) 
at the entrance of the region of interest,
its wavefunction is severely modified (``collapsed''),
so that the times elapsed 
until a further detection when it leaves the region do not reproduce 
the ideal dwell-time operator distribution, and depend on the 
details of the localization method. Proposals for operational, i.e.\ measurement-based approaches to traversal times based on model detectors studying the effect of localization 
have been discussed by Palao et al. \cite{PaMuBrJa-PLA-1997} and by Ruschhaupt \cite{Ru-PLA-1998}. 
All operational approaches to the quantum dwell time known so far provide in fact, and only indirectly, just the average, by deducing it   
from its theoretical relation with some other observable with measurable average.
It is obtained for example by a ``Larmor clock", using a weak homogeneous magnetic field in the region $\R$ and the amount of spin rotations of an incident spin-$\frac{1}{2}$ particle \cite{Baz-SJNP-1967, Ry-SJNP-1967, Bue-PRB-1983}. An optical analogue is provided by the ``Rabi clock" \cite{Bra-JPB-1997}. 
It can also be obtained by measuring the total absorption if a weak complex absorbing potential acts in the region \cite{CP1,CP2,CP3}. This setup could be implemented with 
cold atoms and lasers as described in \cite{CP4}.  

The expression for the average of the dwell time for time-dependent scattering processes 
in terms of the probability densities corresponds (in sharp contrast to its second moment, as shown below) to the classical 
expression for an ensemble of classical particles and it reads \cite{JaWa-PRA-1988, MuBrSa-PLA-1992} 
\beqa
\label{taud}
\tau_D &=& \la \psi|\wh{T}_D|\psi\ra=
\int_{-\infty}^\infty \d t \int_{x_1}^{x_2} \d x\,|\psi(x,t)|^2
\\
&=&\int_0^\infty \d k\,|\la k|{\psi}^{in}\ra|^2 \tau_D(k),
\label{taud1}
\eeqa
where $\psi(x,t) = \int_0^\infty \d k\, \la\phi_k|\psi\ra \exp(-i\hbar k^2 t/2m) \phi_k(x)$ is the time-dependent wave packet and we assume, here and in the rest of the paper, incident 
wavepackets with positive momentum components. 
To write Eq. (\ref{taud1}) use has been made of the standard scattering relation 
$\la \phi_k|\psi\ra=\la k|\psi^{in}\ra$, where  
$\la x|k\ra=(2\pi)^{-{1/2}}\exp(ikx)$, and $\psi^{in}$ is the freely-moving
asymptotic incoming state of $\psi$. 
Integrals of the form (\ref{taud}) had been used
to define time delays by comparing the free motion to that with a scattering center and taking the limit of infinite volume \cite{GoWa-book}. 
         
For a sample of further theoretical studies on the quantum dwell time see
\cite{SK91,BSM94,LM94,Nuss2002,DEMN04,Y04,LV04,W06,K07,S07,BS08}. A recurrent topic has been 
its role and decomposition in tunneling collisions. Instead, we shall focus here
on a different, so far overlooked, but rather fundamental aspect, namely,
the measurability and physical implications of its second moment.           

We shall generalize the approach by Pollak and Miller \cite{PoMi-PRL-1984}, who showed that the average stationary dwell time agrees with the first moment of a microcanonical flux-flux correlation function (ffcf). We demonstrate that this relation holds also for the second moment, and we extend their analysis to the time-dependent (wavepacket) case. The relation fails for third and higher moments and
thus the ffcf contains only part of the information of the dwell-time distribution,  although it is certainly the most relevant. We shall also discuss a possible scheme to measure ffcf's, thus paving the road towards experimental access to quantum 
features of the dwell time distribution.


%
%
%


\section{Stationary flux-flux correlation function}


A connection between the average stationary dwell time and the first moment of  a ffcf has been shown by Pollak and Miller \cite{PoMi-PRL-1984}. They define a quantum microcanonical ffcf $C_{PM}(\tau,k) = \tr\{\Re\, \wh{C}_{PM}(\tau,k) \}$ by means of the operator 
\begin{multline} \label{eq:ffcf_PM}
\wh{C}_{PM}(\tau,k) = 2\pi\hbar [\wh{J}(x_2,\tau) \wh{J}(x_1,0) + \wh{J}(x_1,\tau) \wh{J}(x_2,0)\\ - \wh{J}(x_1,\tau) \wh{J}(x_1,0) - \wh{J}(x_2,\tau) \wh{J}(x_2,0)] \delta(E_k - \wh{H}), 
\end{multline}
where $\wh{J}(x,t) = \e^{i \wh{H} t/\hbar} \frac{1}{2m} [\wh{p} \delta(\wh{x} - x) + \delta(\wh{x} -x) \wh{p}] \e^{-i \wh{H} t/\hbar}$ is the quantum mechanical flux operator in the Heisenberg picture, and $\wh{p}$ and $\wh{x}$  are the momentum and position operators.  

The motivation for this definition stems from classical mechanics and can be understood intuitively: Eq.~(\ref{eq:ffcf_PM}) counts flux correlations of particles entering $\R$ through $x_1$ ($x_2$) and leaving it through $x_2$ ($x_1$) a time $\tau$ later. Moreover, particles may be reflected and may leave the region $\R$ through its entrance point. This is described by the last two terms, where the minus sign compensates for the change of sign of a back-moving flux. Note that these negative terms lead to a self-correlation contribution that diverges for $\tau\to 0$.

We review in the following the derivation of the average correlation time and show afterwards that a similar relation for the second moment holds.
As in the rest of the paper, we shall only consider positive incident momenta, so that we shall actually deal with $\wh{C}^+_{PM}$, substituting $\delta(E_k-\wh{H})$ by  $\delta^+(E_k-\wh{H}):=\delta(E_k-\wh{H})\Lambda_+$, 
where $\Lambda_+$ is the projector onto the subspace of eigenstates of $H$ with positive momentum incidence. 
First of all, we note that, by means of 
the continuity equation,
\be \label{eq:cont}
\frac{\d}{\d x} \wh{J}(x,t) = \frac{\d}{\d t} \wh{\rho}(x,t),
\ee
where $\wh{\rho}(x,t) = \e^{i \wh{H} t/\hbar} \delta(\wh{x} - x) \e^{-i \wh{H} t/\hbar}$ is the (Heisenberg) density operator, 
$\wh{C}^+_{PM}(\tau,k)$ can be written as 
\be \label{eq:C_anders}
\wh{C}^+_{PM}(\tau,k) = - 2\pi\hbar \left( \frac{\d}{\d \tau} \wh{\chi}_\R(\tau) \right) \left( \frac{\d}{\d t} \wh{\chi}_\R(t) \right)_{t=0} \delta^+(E_k - \wh{H}).
\ee
By a partial integration and using the Heisenberg equation of motion the first moment of the Pollak-Miller correlation function is given by
\begin{multline}
\tr \left\{ \int_0^\infty \d \tau\, \tau \wh{C}^+_{PM}(\tau,k) \right\} \\
= \tr \left\{ 2\pi\hbar \int_0^\infty \d\tau \wh{\chi}_\R(\tau) \frac{1}{i\hbar} [\wh{\chi}_\R(0), \wh{H}] \delta^+(E_k - \wh{H})\right\}.
\end{multline}
Boundary terms of the form $\lim_{\tau\to\infty} \tau^{\gamma} \wh{\chi}_\R(\tau), \gamma = 1,2$, are omitted here and in the following. 
This omission can be justified by recalling that physical states must be time-dependent and square-integrable so that the contribution of these terms should vanish when an integration over stationary wavefunctions is performed to account for the wavepacket dynamics. (In the next section we shall discuss explicitly a time dependent version of the correlation function.)  For potential scattering the probability density decays generically as $\tau^{-3}$, which assures a finite 
dwell time average, but for free motion it decays as $\tau^{-1}$ \cite{MDS95},
making 
$\tau_D$ infinite, unless the momentum wave function vanishes at $k=0$
sufficiently fast as $k$ tends to zero \cite{DEMN04}.
  
Writing the commutator explicitly and using the cyclic property of the trace gives
\begin{multline}
\tr \left\{ \int_0^\infty \d \tau\, \tau \wh{C}^+_{PM}(\tau,k) \right\} \\
= \tr \left\{ 2\pi\hbar \int_0^\infty \d\tau\left( -\frac{\d}{\d \tau}\wh{\chi}_\R(\tau) \right) \wh{\chi}_\R(0) \delta^+(E_k - \wh{H})\right\},
\end{multline}
and integration over $\tau$ yields the final result of Pollak and Miller,
\beqa
&&\tr \left\{ \int_0^\infty \d \tau\, \tau \wh{C}^+_{PM}(\tau,k) \right\}
\nonumber\\
&=&2\pi\hbar \tr \left\{ \wh{\chi}_\R(0) \delta^+(E_k - \wh{H}) \right\}
=\mathsf{T}_{kk}.
\eeqa
Expressing the trace in the basis $\ket{\phi_k}$ gives back the stationary dwell time of Eq.~(\ref{eq:dwell_stat}), i.e.\ the diagonal element of the on-the-energy-shell  dwell time operator, $\mathsf{T}_{kk}$.

Note that in our derivation it becomes obvious that the average correlation time is real, in spite of the fact that $\wh{C}^+_{PM}(\tau,k)$ is not selfadjoint. (The discussion of the imaginary part of the ffcf in \cite{PoMi-PRL-1984} is based on a modified version of Eq.~(\ref{eq:ffcf_PM}).)

Next, we will show that the second moment of the Pollak-Miller ffcf equals the second moment of $\mathsf{T}$. This was not observed in Ref.~\cite{PoMi-PRL-1984}. Proceeding in a similar way as above, 
\begin{multline}
\tr \left\{ \int_0^\infty \d \tau\, \tau^2 \wh{C}^+_{PM}(\tau,k) \right\} \\
= \tr \left\{ 4\pi\hbar \int_0^\infty \d\tau\, \wh{\chi}_\R(\tau) \wh{\chi}_\R(0)\delta^+(E_k - \wh{H}) \right\}.
\end{multline}
With $\tr \{\cdots\} = \int \d k' \bra{\phi_{k'}} \cdots \ket{\phi_{k'}}$ 
and taking ``the real part of'' $\wh{C}^+_{PM}$, i.e.\ $[\wh{C}^+_{PM}+(\wh{C}^+_{PM})^\dagger]/2$, we obtain
\beqa
&&\tr \left\{ \Re \int_0^\infty \d \tau\, \tau^2 \wh{C}^+_{PM}(\tau,k) \right\}
=\frac{4\pi^2 m^2}{\hbar^2 k^2}
\nonumber\\
&\times& \left[\left(\int_{x_1}^{x_2} \d x\, |\phi_k(x)|^2 \right)^2\!\!+\left| \int_{x_1}^{x_2} \d x\, \phi_k^*(x) \phi_{-k}(x) \right|^2 \right]
\nonumber\\
&=&(\mathsf{T}^2)_{kk}.
\label{16}
\eeqa
Interestingly, one finds exactly the second moment of $\mathsf{T}$.  
This shows that the relation between dwell times and flux-flux correlation functions goes beyond average values and that $C^+_{PM}(\tau,k)$ includes the quantum features of a dwell time: note that the first summand in Eq. (\ref{16}) is nothing but $(T_{kk})^2$, whereas the second summand is positive, which allows for a non-zero on-the-energy shell dwell-time variance.  We insist that the stationary state considered has positive momentum, $\phi_k(x)$, $k>0$, but this second term  implies the degenerate partner $\phi_{-k}(x)$ as well and is generically non-zero.   

The question arises if these connections hold for the other moments of $C_{PM}^+(\tau,k)$. 
The answer is no, as we will show in the next section with a more
general approach.
%
%
%
%
\section{Time-dependent flux-flux correlation function}
In the following we present a time-dependent version of the above flux-flux correlation function and show its relation to dwell times. So far, ffcf have been mostly considered in chemical physics to define reaction rates for microcanonical or canonical ensembles \cite{MST83}. However, a physically intuitive time-dependent version can be defined in terms of the operator
\begin{multline}
\label{eq:C_op} 
  \widehat{C}(\tau) = \int_{-\infty}^\infty \d t\, \bigl[\widehat{J}(x_2,t+\tau)
  \widehat{J}(x_1,t) + \widehat{J}(x_1,t+\tau)
  \widehat{J}(x_2,t) \\- \widehat{J}(x_1,t+\tau)
  \widehat{J}(x_1,t) - \widehat{J}(x_2,t+\tau)
  \widehat{J}(x_2,t) \bigr],
\end{multline}
which leads to the flux-flux correlation function
\be \label{eq:Ctau_time}
C(\tau) = \langle \Re\,\wh{C}(\tau) \rangle_\psi,
\ee
where the real part is taken to symmetrize the non-selfadjoint operator $\wh{C}(\tau)$
as before.

As in the stationary case, Eq.~(\ref{eq:Ctau_time}) counts flux correlations of particles entering $\R$ through $x_1$ or $x_2$ at a time $t$ and leaving it either through $x_1$ or $x_2$ a time $\tau$ later. Moreover one has to integrate over the entrance time $t$. It is easy to show that the first moment of the classical version of Eq.~(\ref{eq:C_op}), where $\widehat{J}$ is replaced by the classical dynamical variable of the flux, gives the average of the classical dwell time.

As in Eq.~(\ref{eq:C_anders}), we may rewrite $\wh{C}(\tau)$ in the form
\be
\wh{C}(\tau) = -\int_{-\infty}^\infty \d t \frac{\d}{\d \tau} \wh{\chi}_\R(\wh{x},t+\tau) \frac{\d}{\d t} \wh{\chi}_\R(\wh{x},t).
\ee
First of all we note that the ffcf $C(\tau)$ is not normalized, in fact its negative contributions exactly cancel the positive ones, 
\be
\int_0^\infty \d\tau\,\wh{C}(\tau) = \int_{-\infty}^\infty \d t\, \wh{\chi}_\R(t) \frac{\d}{\d t} \wh{\chi}_\R(t) = 0.
\ee
%

Next we derive the average of the time-dependent correlation function. With a partial integration one finds
\begin{multline}
\int_0^\infty \d \tau\,\tau \wh{C}(\tau) = \int_{-\infty}^\infty\! \d t   \int_0^\infty\! \d \tau\, \wh{\chi}_\R(\wh{x},t+\tau)  \frac{\d}{\d t} \wh{\chi}_\R(\wh{x},t).
\end{multline}
A second partial integration with respect to $t$, replacing $\d/\d t$ by $\d/\d \tau$ and integrating over $\tau$ gives
\begin{equation}
\label{eq:result_first}
\int_0^\infty \d \tau\,\tau \wh{C}(\tau) = \int_{-\infty}^\infty\! \d t\, \wh{\chi}_\R^2(\wh{x},t) =\int_{-\infty}^\infty \d t\, \wh{\chi}_\R(\wh{x},t) = 
\wh{T}_D,
\end{equation}
where the projector property of $\wh{\chi}_\R$ has been used. Eq.~(\ref{eq:result_first}) generalizes the result of Pollak and Miller to time-dependent dwell times.

A similar calculation can be performed for the second moment of $C(\tau)$. 
After three partial integrations with vanishing boundary contributions to get rid of the factor $\tau^2$ one obtains
\begin{multline}
\int_0^\infty \d \tau\,\tau^2 \wh{C}(\tau) = 2 \int_{-\infty}^\infty\! \d t \int_0^\infty \d \tau\,\wh{\chi}_\R(\wh{x},t+\tau)\wh{\chi}_\R(\wh{x},t).
\end{multline}
We evaluate the real part of this expression. 
Making the substitutions $t+\tau \to t$ and $\tau \to -\tau$ in the complex conjugated term, we find
\begin{equation} \label{eq:result_second}
\mathrm{Re}\int_0^\infty \d \tau\,\tau^2 \wh{C}(\tau) = \wh{T}_D^2.
\end{equation}
%

\section{Example: free motion}

\begin{figure}
\epsfxsize=8cm
\epsfbox{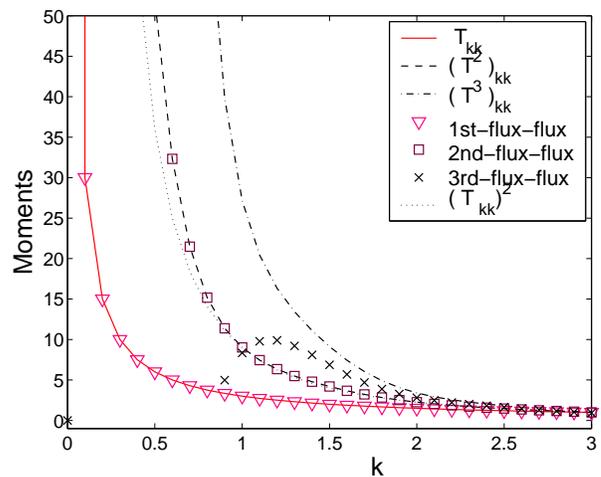}
\caption{Comparison of the first three moments: 
$\mathsf{T}_{kk}$, $(\mathsf{T}^2)_{kk}$ and $(\mathsf{T}^3)_{kk}$ (dotted-dashed line) with the corresponding moments
of the flux-flux correlation function, for a free-motion 
stationary state with fixed $k$. $(\mathsf{T}_{kk})^2$ is also shown (dotted line).  $\hbar = m = 1$ and $L = 3$.}
\label{fig:moments}
\end{figure}

In this section we study the simple case of free motion to show the relations between dwell times and ffcf. For a stationary flux of particles with energy $E_k$, $k>0$, described by $\phi_k(x) = \braket{x}{k} = (2\pi)^{-1/2} \e^{ikx}$, the first three moments of the ideal dwell-time distribution on the energy shell 
are given by
\begin{eqnarray}
\mathsf{T}_{kk} &=& \frac{mL}{\hbar k},\\
(\mathsf{T}^2)_{kk}&=& \frac{m^2L^2}{\hbar^2 k^2} \left(1 + \frac{\sin^2(kL)}{k^2 L^2} \right),
\label{t2}\\
(\mathsf{T}^3)_{kk} &=& \frac{m^3L^3}{\hbar^3 k^3} \left(1 + 3 \frac{\sin^2(kL)}{k^2 L^2} \right).
\end{eqnarray}
As proved above, the first two moments agree with the corresponding moments of the Pollak-Miller ffcf, but for the third moment we obtain
\begin{multline}
\tr\left\{\Re\int_0^\infty \d\tau\, \tau^3 \wh{C}^+_{PM}(\tau,k)\right\}
\\
= \frac{m^3L^3}{\hbar^3 k^3}\left[1 - \frac{3[1+ \cos^2(kL)]}{k^2 L^2} + \frac{3 \sin(2kL)}{L^3 k^3} \right].
\label{28}
\end{multline}
In Fig.~\ref{fig:moments} the first three moments are compared. The agreement between $(\mathsf{T}^3)_{kk}$ and Eq. (\ref{28}) is very
good for large values of $k$, but they clearly differ for small $k$. 

\begin{figure}
\centering
\epsfxsize=8cm  \epsfbox{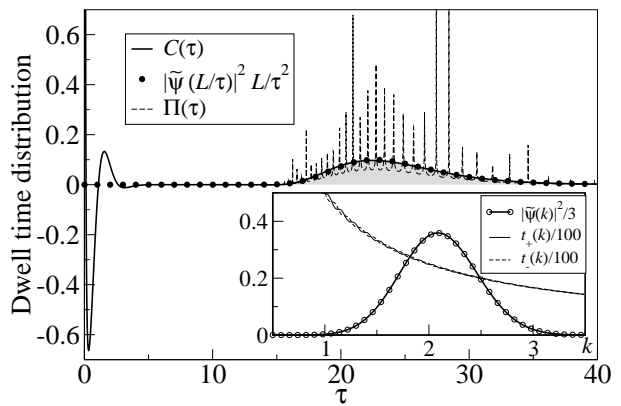}
\caption{Comparison of dwell time distribution $\Pi(\tau)$ (dashed line) and flux-flux correlation function $C(\tau)$ (solid line) for free motion and $L=50$. Furthermore, the alternative free-motion dwell-time distribution $\pi(\tau)= mL|\widetilde{\psi}(mL/(\hbar \tau))|^2/(\hbar \tau^2)$ is plotted (circles). The inset shows the momentum distribution according to Eq.~(\ref{eq:k_distri}) with $k_0=2$, $\Delta k =0.4$, $\alpha = 0.5$ and the eigenvalues $t_\pm(k)$. We set $\hbar=m=1$
and $|x_0|$ large enough to avoid overlap of the initial state
with the space region $\R$.}
\label{fig:Ctau1}
\end{figure}

However, the agreement of the first two moments suggests a similar behavior of $\Pi(\tau)$ and $C(\tau)$. To derive $\Pi(\tau)$, we use the fact that the eigenvalues of $\mathsf{T}$ are $t_\pm(k) = mL[1\pm \sin(kL)/kL]/\hbar k$ and the corresponding eigenstates are $\ket{t_\pm(k)} = (\ket{k}\pm \exp[ik(x_1+x_2)] \ket{-k})/\sqrt{2}$. Then, for a wavefunction $\widetilde{\psi}(k):=\la k|\psi\ra$ with only positive momentum components one obtains from Eq.~(\ref{eq:dwell_distri})
\be \label{eq:dwell_distri_free}
\Pi_D(\tau) = \frac{1}{2} \sum_j \sum_{\gamma = \pm} \frac{|\widetilde{\psi}(k_j^\gamma(\tau))|^2}{|F'_\gamma(k_j^\gamma(\tau))|},
\ee
where $k_j^\gamma(\tau)$ are the solutions of the equation $F_\gamma(k) \equiv t_\gamma(k)-\tau = 0$. In particular, we use the following wavefunction for the calculation \cite{Muga-in-book},
\be \label{eq:k_distri}
\widetilde{\psi}(k) = N(1-\e^{-\alpha k^2})\,\e^{-(k-k_0)^2/[4(\Delta k)^2]} \e^{-ikx_0}\Theta(k),
\ee
where $N$ is the normalization constant and $\Theta(k)$ the step function.
For the free flux-flux correlation function we write
\be
C(\tau) = \Re \int_0^\infty \d k \int_0^\infty \d k'\, \widetilde{\psi}^*(k) \widetilde{\psi}(k') \bra{k} \wh{C}(\tau) \ket{k'},
\ee
and $C_{kk'}(\tau) = \bra{k} \wh{C}(\tau) \ket{k'}$ in the free case is given by
\beqa
C_{kk'}(\tau) &=& \frac{m}{2\pi \hbar k} \delta(k-k') \frac{\d^2}{\d\tau^2} \Big[2f(\hbar k\tau/m) - f(\hbar k\tau/m-L)
\nonumber\\
&-& f(\hbar k\tau/m+L)\Big],
\eeqa
where 
\be
f(x) = -2\e^{imx^2/(2\hbar\tau)} \left(\frac{i\pi\hbar\tau}{2m} \right)^{1/2} + i\pi x\, \mt{erfi} \left(\sqrt{\frac{im}{2\hbar \tau}}\,x \right).
\ee
The result is shown in Fig.~\ref{fig:Ctau1}. The ffcf shows a hump around the mean dwell time, as expected (the area under this hump is $0.9993$), but it strongly oscillates for small $\tau$ and diverges for $\tau\to 0$. As discussed above, this is due to the self-correlation contribution of wavepackets which are at $x_1$ or $x_2$ at the times $t$ and $t+\tau$ {\it without} changing the direction of motion in between. A similar feature has been observed in a traversal-time distribution derived by means of a path integral approach \cite{Fertig-PRL-1990}. 

In contrast, $\Pi(\tau)$ behaves regularly for $\tau\to 0$, but shows peaks in the region of the hump. This is because the denominator of Eq.~(\ref{eq:dwell_distri_free}) becomes zero if the slope of the eigenvalues $t_\pm(k)$ is zero, which occurs at every crossing
point of $t_+(k)$ and $t_-(k)$.

Another dwell time distribution for free motion can be defined based on the operator $\wh{t}_D = mL/|\wh{p}|$, obtained heuristically from quantization of the classical dwell time. The eigenfunctions of this operator are momentum eigenfunctions, $\ket{\pm k}$, $k>0$, and the corresponding eigenvalues are twofold degenerate and equal to the classical time, $mL/\hbar|k|$. The distribution of dwell times for this operator, as always for
positive-momentum states, is given by
\beq
\pi(\tau) = \frac{mL}{\hbar \tau^2}  \left| \widetilde{\psi}\left(\frac{mL}{\hbar\tau}\right) \right|^2.
\eeq
The distribution $\pi(\tau)$ agrees with $C(\tau)$ in the region of the dwell-time peak and tends to zero for $\tau\to 0$. However, it does not show the resonance peaks of $\Pi(\tau)$.
The on-the-energy-shell version of $\widehat{t}_D$, $\mathsf{t}$, is also worth examining. By factoring out an energy delta function 
as in Eq. (\ref{dele}) we get for a plane wave $|k\ra$ the average 
$\mathsf{t}_{kk}=mL/(\hbar k)$, which is equal to $\mathsf{T}_{kk}$,  but the second moment 
differs, $(\mathsf{t}^2)_{kk}=(\mathsf{t}_{kk})^2=(\mathsf{T}_{kk})^2\le (\mathsf{T}^2)_{kk}$, see Fig. 1;
in other words, the
variance on the energy shell is zero since only one eigenvalue is possible for 
$\mathsf{t}$.
Contrast this with the extra term in Eq. (\ref{t2}) which again emphasizes the   
non-classicality of the dwell-time operator
$\widehat{T}_D$ and its quantum fluctuation.    
\section{Discussion}
\begin{figure}
\epsfxsize=8cm
\epsfbox{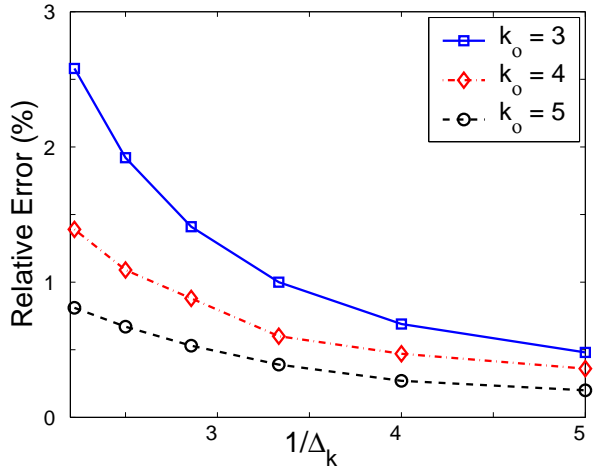}
\caption{Comparison of the relative error of $\langle \wh{T}_D\rangle$ using the approximation ${C_{0}} (\tau)$ instead of $C(\tau)$ for free motion.
$\alpha=0.5$, $\hbar = m = 1$ and $L = 100$.}
\label{fig:delta}
\end{figure}

We have demonstrated that the relationship found by Pollak and Miller \cite{PoMi-PRL-1984} between the first moment of a distribution ffcf and the average stationary dwell time is also valid for the second moment and for flux-flux correlations of wavepackets. On the other hand, this relationship is not valid for the third moment. 
While this brings dwell-time information closer to experimental realization, the difficulty is translated onto 
the measurement of the ffcf, which is not necessarily an easy task. 
The simplest approximation is to substitute the expectation of the 
product of two flux operators by the product of their  
expectation values (the product of the current densities). 
Using the wave packet of Eq. (\ref{eq:k_distri}), we have compared the times obtained 
with the full expression (\ref{eq:Ctau_time}) 
and with this approximation in Fig. \ref{fig:delta}. 
The two results approach as $\Delta_k \to 0$.  
%
%
%
Other factors that improve the approximation are the increase of $L$
and/or of  $k_0$.

This approximation can in fact be improved systematically, still making 
use of ordinary current densities, as follows:   
First we decompose  
$\widehat{J}(x_i,t+\tau)\widehat{J}(x_j,t)$  
by means of the resolution of the identity 
\beqa \label{eq:unit}
\widehat{1}&=&\widehat{P}+\widehat{Q},
\\
\widehat{P}&=&\ket{\psi}\bra{\psi},
\eeqa
%
%
%
so that 
%
%
%
\be 
\label{eq:jj}
\widehat{J}(x_i,t+\tau)\widehat{J}(x_j,t)=\widehat{J}(x_i,t+\tau)(\ket{\psi}\bra{\psi}+\widehat{Q})\widehat{J}(x_j,t).
\ee
It is useful to decompose $\widehat{Q}$ further in terms of a basis of states 
orthogonal to $\ket{\psi}$ and to each other, $\{\ket{\psi_j^Q}\}$, 
%
%
%
\be
\label{eq:q_explic}
\widehat{Q}=\sum_j\ket{\psi_j^Q}\bra{\psi_j^Q},
\ee
%
%
%
that could be generated systematically e.g., by means of a Gram-Schmidt orthogonalization. 
Now we can split Eq. ({\ref{eq:C_op}) as follows 
\be\label{eq:C_op_1y2} 
\widehat{C}(\tau) = \widehat{C}_0(\tau)+\widehat{C}_1(\tau), 
\ee
where 
\beqa
\label{eq:C_op_1} 
\widehat{C}_0(\tau)&=&\int_{-\infty}^\infty \d t\, \bigl[\widehat{J}(x_2,t+\tau)\ket{\psi}\bra{\psi}\widehat{J}(x_1,t) 
\nonumber\\
&+&  \widehat{J}(x_1,t+\tau)\ket{\psi}\bra{\psi}\widehat{J}(x_2,t) 
\nonumber\\
&-& \widehat{J}(x_1,t+\tau)\ket{\psi}\bra{\psi}\widehat{J}(x_1,t) 
\nonumber\\
&-& 
\widehat{J}(x_2,t+\tau)\ket{\psi}\bra{\psi}\widehat{J}(x_2,t) \bigr],
\eeqa
\beqa
\label{eq:C_op_2} 
\widehat{C}_1(\tau)&=&\sum_j\int_{-\infty}^\infty \d t\, \bigl[\widehat{J}(x_2,t+\tau)
 	\ket{\psi_j^Q}\bra{\psi_j^Q}\widehat{J}(x_1,t)
\nonumber\\
&+& \widehat{J}(x_1,t+\tau)
  \ket{\psi_j^Q}\bra{\psi_j^Q}\widehat{J}(x_2,t) 
\nonumber\\
&-&\widehat{J}(x_1,t+\tau)
 	\ket{\psi_j^Q}\bra{\psi_j^Q}\widehat{J}(x_1,t)
\nonumber\\
&-&\widehat{J}(x_2,t+\tau)
\ket{\psi_j^Q}\bra{\psi_j^Q}\widehat{J}(x_2,t) \bigr]. 
\eeqa
Similarly, we define 
$C(\tau)=C_0(\tau)+C_1(\tau)$ by taking the real 
real part of $\la \psi|\wh{C}_0(\tau)+\wh{C}_1(\tau)|\psi\ra$. 
$C_0$ is the zeroth order approximation discussed before and only involves 
ordinary, 
measurable current densities \cite{DEHM2002}.  
The non-diagonal terms from $C_1$,  
$\bra{\psi}\widehat{J}(x_i,t)\ket{\psi_j^Q}\bra{\psi_j^Q}\widehat{J}(x_j,t+\tau)\ket{\psi}$	can also be related to ordinary fluxes by means of the  
auxiliary states
\beqa
\ket{\psi_1}&=&\ket{\psi}+\ket{\psi_j^Q},
\nonumber
\\
\ket{\psi_2}&=&\ket{\psi}+i\ket{\psi_j^Q},
\nonumber
\\
\ket{\psi_3}&=&\ket{\psi}-i\ket{\psi_j^Q},  
\eeqa
since one easily finds that 
\beqa
\label{eq:pol_met}
\bra{\psi}\widehat{J}(x,t)\ket{\psi_j^Q}&=&\frac{1}{2}\bra{\psi_1}\widehat{J}(x,t)\ket{\psi_1}
-\frac{1}{4}\bra{\psi_2}\widehat{J}(x,t)\ket{\psi_2}
\nonumber\\
&-&\frac{1}{4}\bra{\psi_3}\widehat{J}(x,t)\ket{\psi_3}
+\frac{i}{4}\bra{\psi_3}\widehat{J}(x,t)\ket{\psi_3}
\nonumber\\
&-&\frac{i}{4}\bra{\psi_2}\widehat{J}(x,t)\ket{\psi_2}.
\eeqa

To summarize, the present paper provides a route of access to the 
second moment of the quantum dwell time through flux-flux correlation functions. 
This is interesting because the second moment is characteristically quantum
and, unlike the first moment, it differs structurally from (and is larger than) the 
corresponding classical quantity: on the energy shell the dwell time shows a 
quantum fluctuation (non-zero variance) which vanishes classically.  
While the analysis of the 
quantum dwell time has been mostly limited to its average in the existing studies,
the present results 
motivate further research on the role played by the second moment
of the dwell time in fields such as lifetime fluctuations, chaotic systems, conductivity, or time-frequency metrology. For example, is the second moment 
affecting the quality factor of atomic clocks? The time spent by the 
atom in a spatial region  
determines their stability, which increases in principle for slower atoms, but  quantum motion effects have been shown to become  
more and more relevant for decreasing velocities \cite{sei07,mou07}.  
This an other intriguing questions on the quantum dwell time 
are left for separate studies.

\begin{acknowledgments}
We acknowledge discussions  with I. Egusquiza, J. A. Damborenea and B. Navarro. 
This work has been supported by Ministerio de Educaci\'on y Ciencia (FIS2006-10268-C03-01) and the Basque Country University (UPV-EHU, GIU07/40). 
%
%
\end{acknowledgments}


\begin{thebibliography}{99}

\bibitem{Muga-book}
J. G. Muga, R. Sala Mayato, I. L. Egusquiza (eds.), {\it Time in Quantum Mechanics - Vol. 1}, (Springer, Berlin, 2008).

\bibitem{MuLea-PR-2000}
J. G. Muga, G. R. Leavens, Phys. Rep. {\bf 338}, 353 (2000).

\bibitem{DEHM2002}
J. A. Damborenea, I. L. Egusquiza, G. C. Hegerfeldt and J. G. Muga, 
{\it Phys. Rev.} A {\bf 66}, 052104 (2002).
\bibitem{Smith60}
F. Smith, Phys. Rev. {\bf 118}, 349 (1960).
\bibitem{Nuss2002} C. A. A. de Carvalho and H. M. Nussenzveig, Phys. Rep. {\bf 364}, 83 (2002).


\bibitem{HauSto-RMP-1989}
E. H. Hauge, J. A. St{\o}vneng, Rev. Mod. Phys. {\bf 61}, 917 (1989).

\bibitem{LaMa-RMP-1994}
R. Landauer, T. Martin, Rev. Mod. Phys. {\bf 66}, 217 (1994).

\bibitem{EkSie-AP-1971}
H.~Ekstein and A. J. F.~Siegert, Ann. Phys. {\bf 68}, 509 (1971).

\bibitem{JaWa-PRA-1989}
W. Jaworski, D. M. Wardlaw, Phys. Rev. A {\bf 40}, 6210 (1989).

\bibitem{Bue-PRB-1983}
M. B\"uttiker, Phys. Rev. B {\bf 27}, 6178 (1983).

\bibitem{PaMuBrJa-PLA-1997}
J. P. Palao, J. G. Muga, S. Brouard, A. Jadczyk, Phys. Lett. A {\bf 233}, 227 (1997).

\bibitem{Ru-PLA-1998}
A. Ruschhaupt, Phys. Lett. A {\bf 250}, 249 (1998).

%
%
%

\bibitem{Baz-SJNP-1967}
A. Baz', Sov. J. Nucl. Phys. {\bf 4}, 182 (1967).

\bibitem{Ry-SJNP-1967}
V. Rybachenko, Sov. J. Nucl. Phys. {\bf 5}, 635 (1967).

\bibitem{Bra-JPB-1997}
C. Bracher, J. Phys. B: At. Mol. Opt. Phys. {\bf 30}, 2717 (1997).


\bibitem{CP1} R. Golub, S. Felber, R. G\"ahler, and E. Gutsmiedl, Phys.
Lett. A {\bf 148}, 27 (1990).
\bibitem{CP2} Z. Huang and C. M. Wang, J. Phys.: Cond. Matt {\bf 3}, 5915
(1991).
\bibitem{CP3} J. G. Muga, S. Brouard, and R. Sala, J. Phys.: Cond.
Matt. {\bf 4}, L579 (1992).
\bibitem{CP4} A. Ruschhaupt, J. A. Damborenea, B. Navarro,
J. G. Muga and G. C. Hegerfeldt, Europhys. Lett. {\bf 67}, 1 (2004).
\bibitem{JaWa-PRA-1988}
W. Jaworski, D. M. Wardlaw, Phys. Rev. A {\bf 37}, 2843 (1988).
\bibitem{MuBrSa-PLA-1992}
J.G. Muga, S. Brouard, R. Sala, Phys. Lett. A {\bf 167}, 24 (1992).
\bibitem{GoWa-book}
M.L. Goldberger, K.M. Watson, {\it Collision theory} (Krieger, Huntington 1975)
\bibitem{SK91} D. Sokolovski and J. N. L. Connor, Phys. Rev. A {\bf 44}, 1500 (1991).
\bibitem{BSM94} S. Brouard, R. Sala, and J. G. Muga, Phys. Rev. A {\bf 49}, 4312
(1994). 
\bibitem{LM94} C. R. Leavens, W. R. McKinnon, Phys. Lett. A {\bf 194}, 12 (1994).
\bibitem{DEMN04}
J.A. Damborenea, I.L. Egusquiza, J.G. Muga, B. Navarro, arXiv:quant-ph/0403081 (2004). 
\bibitem{Y04} N. Yamada, Phys. Rev. Lett. {\bf 93}, 170401 (2004).
\bibitem{LV04} H. Lewenkopf and R. O. Vallejos, Phys. Rev. E {\bf 70}, 036214 (2004).
\bibitem{W06} H. G. Winful, Phys. Rep. {\bf 436}, 1 (2006). 
\bibitem{K07} N. G. Kelkar, Phys. Rev. Lett. {\bf  99}, 210403 (2007).
\bibitem{S07} D. Sokolovski, Phys. Rev. A {\bf 76}, 042125 (2007).
\bibitem{BS08} S. Boonchui and V. Sa-yakanit, Phys. Rev. {\bf 77}, 044101 (2008).
\bibitem{PoMi-PRL-1984}
E. Pollak and W. H. Miller, Phys. Rev. Lett. {\bf 53}, 115 (1984).
\bibitem{MDS95} J. G. Muga, V. Delgado, and R. F. Snider, Phys. Rev. {\bf 52},
16381 (1995).
\bibitem{MST83} W. H. Miller, S. D. Schwartz, and J. W. Tromp, J. Chem. Phys. {\bf 79}, 4889 (1983).
\bibitem{Muga-in-book}
J.G. Muga, {\it Characteristic Times in One-Dimensional Scattering}, in: J.G. Muga, R. Sala Mayato, I.L. Egusquiza (eds.), {\it Time in Quantum Mechanics}, (Springer, Berlin, 2002), pp. 29-68.
\bibitem{Fertig-PRL-1990}
H.A. Fertig, Phys. Rev. Lett. {\bf 65}, 2321 (1990).
\bibitem{sei07} D. Seidel and J. G. Muga, Eur. Phys. J. D {\bf 41}, 71 (2007).
\bibitem{mou07} S. V. Mousavi, A. del Campo, I. Lizuain, and J. G. Muga, 
Phys. Rev. A {\bf 76}, 033607 (2007). 

%

\end{thebibliography}
\end{document}